\documentclass[aps,prl,preprint,superscriptaddress]{revtex4}

\usepackage{amsmath}
\usepackage{amssymb}
\usepackage{graphicx}
\usepackage{bm}

\begin{document}

\title{Polarization proximity effect in isolator crystal pairs}

\author{Y. Linzon}
\affiliation{Universite du Quebec, Institute National de la
Recherche Scientifique, Varennes, Quebec J3X 1S2, Canada}

\author{M. Ferrera}
\affiliation{Universit\'{e} du Qu\'{e}bec, Institut\'{e} National de
la Recherche Scientifique, Varennes, Qu\'{e}bec J3X 1S2, Canada}

\author{L. Razzari}
\affiliation{Universit\'{e} du Qu\'{e}bec, Institut\'{e} National de
la Recherche Scientifique, Varennes, Qu\'{e}bec J3X 1S2, Canada}

\author{A. Pignolet}
\affiliation{Universit\'{e} du Qu\'{e}bec, Institut\'{e} National de
la Recherche Scientifique, Varennes, Qu\'{e}bec J3X 1S2, Canada}

\author{R. Morandotti}
\affiliation{Universit\'{e} du Qu\'{e}bec, Institut\'{e} National de
la Recherche Scientifique, Varennes, Qu\'{e}bec J3X 1S2, Canada}

\begin{abstract}We experimentally studied the polarization dynamics (orientation and ellipticity) of near infrared light transmitted through magnetooptic Yttrium Iron Garnet crystal pairs using a modified balanced detection scheme. When the pair separation is in the sub-millimeter range, we observed a proximity effect in which the saturation field is reduced by up to 20\%. 1D magnetostatic calculations suggest that the proximity effect originates from magnetostatic interactions between the dipole moments of the isolator crystals. This substantial reduction of the saturation field is potentially useful for the realization of low-power integrated magneto-optical devices.\end{abstract}



\maketitle

\noindent Optical isolators are important polarization components
which are controlled by static external magnetic fields. They induce
nonreciprocal polarization phase shifts originating from the
so-called magneto-optical (MO) Faraday rotation [1]. The
unidirectional nonreciprocal polarization control is crucial for the
reduction of reflection-related instabilities in active devices
[1-6]. A fully functional isolator consists of a magnetically active
isolator crystal situated between two crossed polarizers, and a
rotation equal to $\pm$45$^{\circ}$ of an input linear polarization
is required at the wavelength of operation, in the presence of an
external magnetic field $\pm$$H$. Ferrimagnetic iron garnets are
popular materials for isolators in the visible and near infrared, as
they possess large induced magnetizations, leading to the highest
known Faraday rotations in the spectral range used for today's
optical telecommunication systems [7]. Specifically, in a Yttrium
Iron Garnet (YIG, Y$_{3}$Fe$_{5}$O$_{12}$) crystal subject to a
magnetic field, the latent Faraday angle per unit thickness, at the
principal telecom wavelength (1.55 $\mu$m), is typically  $\sim$
0.016$^{\circ}/\mu \text{m}$ at saturation, when the light
propagates along the crystal's easy axis of magnetization [7,8]. As
a result, a YIG crystal operating in the telecom spectral region
typically requires a length of several millimeters to be effective
as an optical isolator. In propagation of light through such thick
MO samples, the (usually) desirable circular birefringence is
accompanied by substantial magnetically-induced circular dichroism
[8], which introduces an ellipticity ("smearing") $\varepsilon$ of
the polarization [9].

While the optical response of MO single crystals as a function of
the magnetic field has been thoroughly characterized in the past
[2-8], the polarization dynamics associated to the combination of
several (i.e., separate) MO components, located in close proximity
and subject to a uniform magnetic field, has not been addressed so
far. Such a study, however, is important in view of recent potential
applications involving the integration of several optoelectronic
devices on the same chip [2-6,10]. In such devices the overall
response of a set of magneto-optical components, including
magnetostatic proximity effects [11], plays a significant role and
can lead to unexpected and surprising behaviors. In this paper we
report our study of the polarization dynamics (in terms of
orientation and ellipticity, as a function of the magnetic field) of
light transmitted through YIG isolator crystal pairs positioned in
series (i.e. one after the other), using a balanced detection
scheme. For small separations of the YIG crystal pairs, we observed
a magnetostatic proximity effect [11], in which the saturation field
is reduced significantly, i.e. by up to 20\%.

\begin{figure}[b]
\centerline{\includegraphics{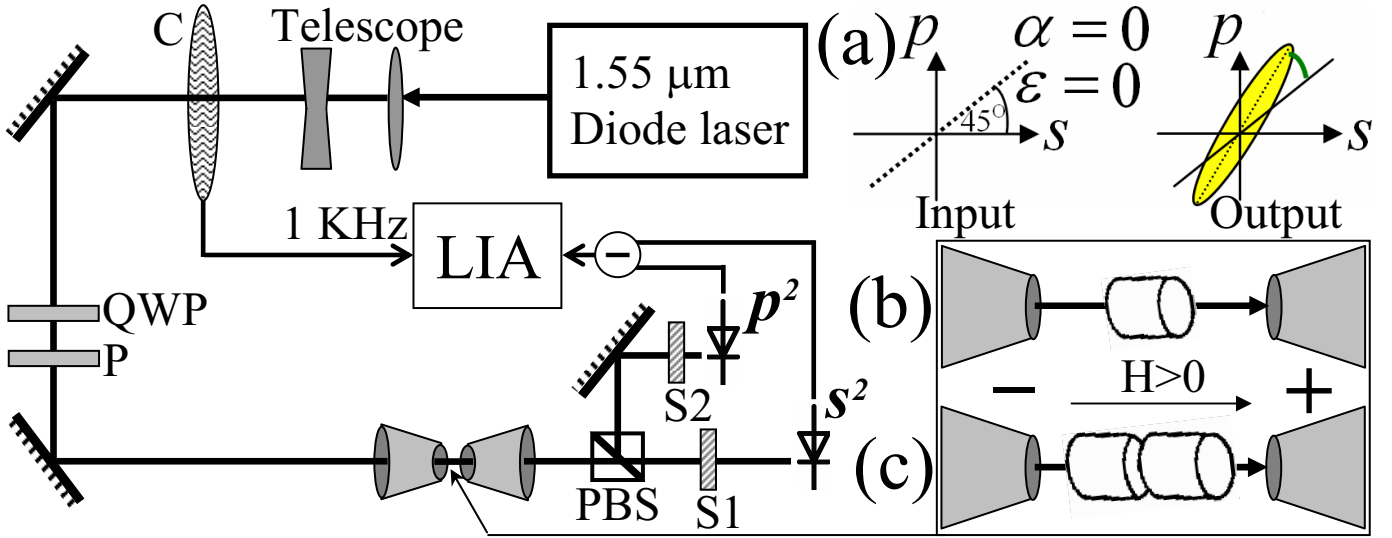}} \caption{(Color online)
Experimental schematics, as described in the text. (a) Definition of
input (left) and output (right) polarizations. (b)-(c) Volume within
the magnet with (b) single and (c) double crystal settings.}
\end{figure}

The experimental setup is represented in Fig. 1. A linearly
polarized diode laser beam at a wavelength of 1.55 $\mu$m is
collimated with a telescope and passed through a chopper (C) which
synchronizes a lock-in amplifier (LIA) at 1 KHz. The combination of
a quarter wave-plate (QWP) and a polarizer (P) enables the control
of the input linear polarization. The input polarization is always
set initially to a value of 45$^{\circ}$ in the first quadrant
[$s>$0 and $p>$0, see Fig. 1(a)]. After propagation along the
magnet, the output beam enters a polarizing beam splitter (PBS)
which separates the $s$ and $p$ components, both of which are
detected by a pair of identical photodiode detectors connected to
the lock-in amplifier operating in differential mode. Since optical
detectors always measure intensity rather than amplitude, the signal
read by the amplifier when both channels are open is $s^2-p^2$. When
either one or both components change sign, a vectorial correction
must be applied, as described below. Each polarization component can
be blocked individually with a mechanical shutter (S1 or S2) to read
the other component independently. The total optical energy is
proportional to $s^2+p^2$, a constant quantity throughout all scans
of the magnetic field. In our experiments, we used a GMW magnet
(model 3470). With a separation of 2 cm between the poles, the
magnet delivers a nearly uniform magnetic field of up to 4 kG within
the volume between the poles. Inside the magnet, either a single
crystal [Fig. 1(b)] or a pair of crystals [Fig. 1(c)] have been
used. Each MO sample is a 2.8 mm-long single crystal YIG rod of 5 mm
diameter, polished and anti-reflection coated for operation at the
laser's wavelength. The easy axis of magnetization is nearly
parallel to the rod axis, which is also the optical beam and
magnetic field directions [see Figs. 1(b),(c)]. The transmission is
88\% with a single crystal and 80\% with the pair setting.

A general output state of polarization has both an orientation angle
$\alpha$ with respect to the input state, and an ellipticity angle
$\epsilon$ [9]. We define a normalized balanced detection signal as
$D=(s^2-p^2)/(s^2+p^2)$. In the presence of ellipticity the entire
energy can never be set parallel to a single linear polarization
state (as opposed to the input state): some residual energy always
remains in the other orthogonal state, corresponding to the minor
axis of the ellipse. Closing S1 and maximizing the signal by
rotating P yields a signal $p^2_{max}$ that corresponds to alignment
of the ellipse's major axis along $p$. In the same setting, closing
S2 and opening S1 yields a corresponding signal $s^2_{min}$ relating
to the ellipse's minor axis aligned along $s$. The square root of
the ratio between these signals is thus proportional to the ratio
between the polarization ellipse axes, and we can thus define a
normalized signal $\eta=\sqrt{(|p^2_{max}|)/(|s^2_{min}|)}$ which is
measured in conjunction with $D$. A straightforward calculation
yields the following relation between the measured signals, $D$ and
$\eta$, and the physical parameters $\alpha$ and $\epsilon$:
\begin{align}
D\cdot(\eta^{2}+1)=(\eta^{2} - 1) \cdot sin(2\alpha) \mp 2\eta cos(2\alpha),\nonumber\\
\varepsilon=arctan(\eta),
\end{align}
The sign in the right-hand term of the first equation corresponds to
different quadrants, in the $s-p$ plane, of the output polarization.
The actual quadrant of the polarization must thus be taken into
account before applying Eq. (1). Each equation yields two real
solutions, of which only one has physical meaning. This measurement
is sensitive to the sign of $\alpha$, but not to that of $\epsilon$.

\begin{figure}[t]
\centerline{\includegraphics[width=12cm]{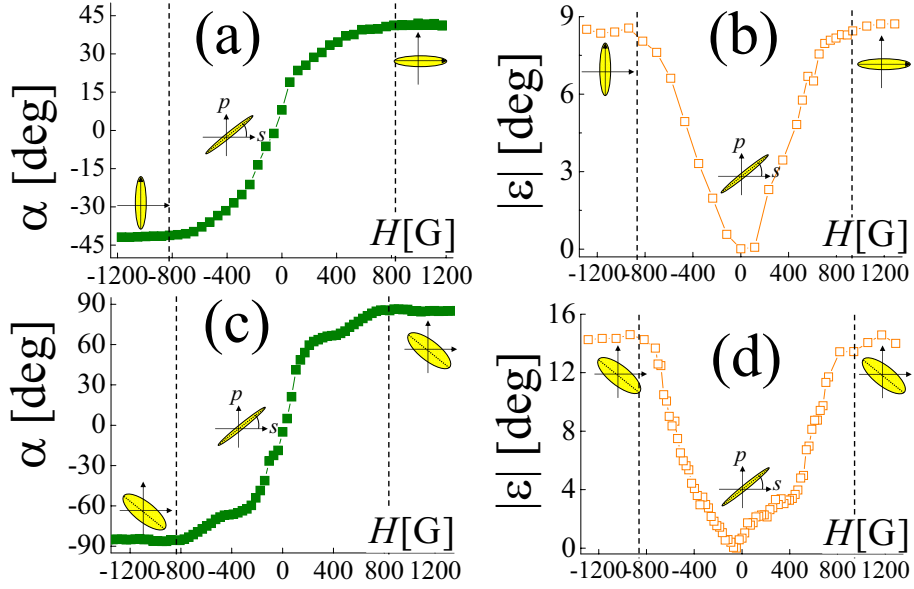}}
\caption{(Color online) Polarization dynamics as a function of the
magnetic field in the single crystal [(a),(b)] and in distant
crystal pair [(c),(d)] geometries. The polarization parameters
$\alpha$ [(a),(c)] and $\varepsilon$ [(b),(d)] are in degrees.}
\end{figure}

Experimental results are shown in Fig. 2. Considering first the
single crystal geometry [Fig. 1(b)], the polarization dynamics
(angles $\alpha$ and $\epsilon$ calculated from Eq. (1), as a
function of the magnetic field), are shown in Figs. 2(a),(b). Here
the polarization is always in the first quadrant of the $s-p$ plane,
corresponding to the minus sign in Eq. (1). The slopes obtained from
linear fits to $\alpha(H)$ and $\varepsilon(H)$, in fields far below
saturation, are 4.1$\times$10$^{-2}$ $^{\circ}/(\text{G$\times$mm})$
for the Faraday rotation (Verdet coefficient) and
4.5$\times$10$^{-3}$ $^\circ{}/(\text{G$\times$mm})$ for the
magnetic circular dichroism, in accordance with the literature
[7,8]. The orientation and ellipticity both saturate in a magnetic
field of $\pm$820 G at the values of $\sim\pm$45$^{\circ}$ and
8$^{\circ}$, respectively. Turning to the double-crystal geometry
[Fig. 1(c)], with a 1 mm separation between the crystals, the
polarization dynamics are shown in Figs. 2(c),(d). Above half the
saturation field, the polarization is in the second or fourth
quadrant, and the plus sign is used in Eq. (1). The slopes of
$\alpha(H)$ and $\varepsilon(H)$ far below saturation are
approximately twice those obtained in the single crystal case,
implying that the Verdet and the dichroism coefficients, normalized
by the total thickness, are equal. The orientation and ellipticity
again saturate at $\pm$820 G, and their values are $\sim \pm
$90$^{\circ}$ and 15$^{\circ}$, respectively.

As the distance between the crystals becomes small, however, we
observe a strong dependence of the saturation field of both the
orientation and ellipticity angles on the crystal separation. Figure
3(a) shows the dynamics of $\varepsilon$ for distant (squares) and
adjacent (circles) crystal pair settings. While the saturation field
is $\pm$820 G when the two crystals are set apart, similar to the
case of a single crystal, in adjacent crystal settings the
saturation field is reduced to $\pm$650 G, namely to 80\% of the
initial value. A detailed characterization of the saturation field
($H_{s}$) as a function of the pair facet-to-facet separation ($d$)
[Fig. 3(b)] reveals that the proximity effect decays on a sub-mm
separation scale. As both the Faraday rotation and the magnetic
circular dichroism are proportional to the total magnetization, this
exponential dependence of the decay on the distance between the two
crystals suggests that the proximity effect originates from
effective magnetostatic interactions between the crystal magnetic
moments in the presence of an external field [11].

\begin{figure}[t]
\centerline{\includegraphics[width=12cm]{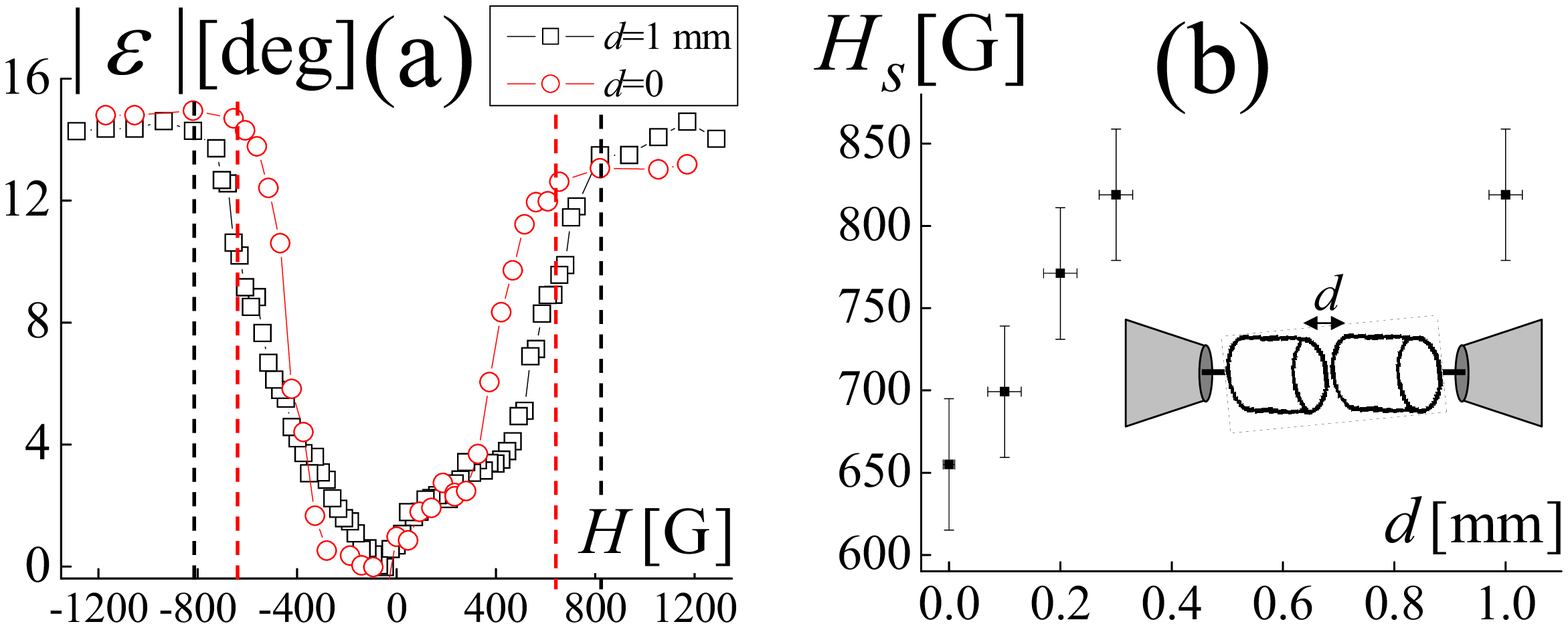}}
\caption{(Color online) Effect of the isolator proximity on the
saturation field. (a) Ellipticity dynamics in far ($d$=1 mm,
squares) and near ($d$=0, circles) pair settings. (b) Saturation
field $H_{s}$ (in terms of both ellipticity and orientation) as a
function of the separation $d$.}
\end{figure}

To further support our interpretation of the proximity effect we
used a simple 1D theoretical model, the results of which are shown
in Fig. 4. A magnetic dipole moment, dispersed around the center,
can be assigned to each isolator crystal [see Fig. 4(a)]. In
saturation, the typical magnetization along the easy axis is $M_{s}=
$140 $\text{emu}/\text{cm}^{3}$ in YIG [7,8]. Integration along the
crystal volume yields a magnetic moment of $m_{s}= $77
$\text{A}\cdot\text{cm}^{2}$. In the presence of an external
magnetic field $H_{ext}=f \cdot H_{s}$, where $0<f<1$ and $H_s$=820
G, the magnetic moment distribution within each crystal is taken as
$m(x)=f \cdot m_{s} h(x)$, where $h(x)$ is the density per unit
length of the dipole distribution along $x$. Here $h$ is taken as a
Gaussian distribution with a FWHM of 2.8 mm (the isolator length)
and an area of unity. As the easy axis of magnetization, the
magnetic field and the beam axis are all approximately parallel to
$x$, we assume that both dipoles are oriented along $x$ and consider
only the dependence on this dimension. Each individual crystal
induces the magnetic field of a dipole dispersed around its center,
$H_{dipole}(x)=\frac{\mu_{0}}{2 \pi}
\int\frac{m(x')\hat{x}}{(x-x')^{3}}dx'$. Thus, the effective
proximity field increase experienced by the right crystal due to the
left crystal is, to the first order,
$H_{prox.}=\int{H_{dipole(left)}(x) \cdot h_{(right)}(x)dx}$ [see
Fig. 4(a)]. This is the same as the effective field induced on the
left crystal by the right crystal, following symmetry reasons. In
addition, the field increase required to reach saturation in each
crystal is $H_{req.}=H_{s}-H_{ext}$. Figs. 4(b)-(d) show the
calculated fields $H_{prox.}$ and $H_{req.}$ as a function of the
crystal separation $d$ for different values of $f$. For low values
of $f$, $H_{prox.}$ and $H_{req.}$ do not coincide for any value of
$d$ [Fig. 4(b)]. This implies that effective saturation cannot be
reached as a result of the proximity effect. For higher values of
$f$ [Figs. 4(c),(d)], $H_{prox.}$ and $H_{req.}$ intersect. In
particular, for separations smaller than $d^{*}$ [indicated in Figs.
4(c),(d)] each crystal is effectively saturated, and $d^{*}$ can
thus be regarded as an effective "saturating" separation. Figure
4(e) shows the field required to reach effective saturation,
$H_{s(eff)}=H_{s}-H_{req.}$, versus $d^{*}$, as extrapolated from
different values of $f$. This can be directly compared to the
experimental result in Fig. 3(b). The effective decrease of the
saturation field by $\sim$20\% in adjacent crystals, as well as an
exponential decay with the distance, are indeed verified [Fig.
4(e)].

\begin{figure}[t]
\centerline{\includegraphics[width=12cm]{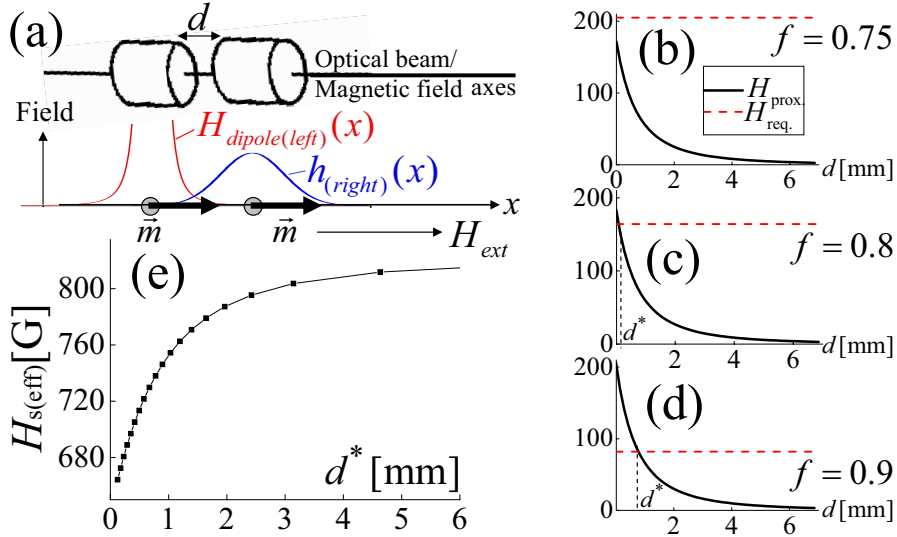}}
\caption{(Color online) 1D calculation of the magnetostatic
proximity effect. (a) Schematic of the model, as described in the
text. (b)-(d) Calculated proximity field (solid line) and required
field increase (dashed line) as a function of the separation $d$, in
external fields corresponding to: (b) $f$=0.75, (c) $f$=0.8, (d)
$f$=0.9. (e) External field required for effective crystal
saturation as a function of the saturating separation $d^{*}$.}
\end{figure}

In conclusion, we have studied the polarization dynamics in YIG
isolator crystal pairs using a modified balanced detection setup. A
proximity effect was observed in which the saturation field is
reduced by up to 20\% for nearby crystals. The decay rate suggests
that this effect originates from magnetostatic interactions between
the nearby magnetic moments in the presence of an external field
[11], as further supported by 1D magnetostatic calculations. This
substantial reduction of the saturation field is potentially useful
for the realization of integrated magneto-optical devices [2-6,10],
since a lower saturating field would essentially mean lower power
consumption.

This research was supported by the Natural Sciences and Engineering
Research Council (NSERC) of Canada. Y.L. acknowledges a MELS FQRNT
fellowship. We also thank R. Helsten for his invaluable technical
assistance.


\begin{thebibliography}{99}

\bibitem{Dillon1} "Magnetooptics", J. F. Dillon, J. Magn. Magn. Mater. {\bf 100}, 425 (1991).

\bibitem{isolator1} "Lens-free in-line optical isolators", T. Sato, J. Sun, R. Kasahara, and S. Kawakami, Opt. Lett. {\bf 24}, 1337 (1999).

\bibitem{isolator2} "Magneto-optical nonreciprocal phase shift in garnet/silicon-on-insulator waveguides", R. L. Espinola, T. Izuhara, M. C. Tsai, and R. M. Osgood, Opt. Lett. {\bf 29}, 941 (2004).

\bibitem{isolator3} "Polarization rotation enhancement and scattering mechanisms in waveguide magnetophotonic crystals", M. Levy and R. Li, Appl. Phys. Lett. {\bf 89}, 121113 (2006).

\bibitem{isolator4} "Demonstration of quasi-phase-matched nonreciprocal polarization rotation in III-V semiconductor waveguides incorporating magneto-optic upper claddings", B. M. Holmes and D. C. Hutchings, Appl. Phys. Lett. {\bf 88}, 061116 (2006).

\bibitem{isolator5} "Magneto-optical isolator with silicon waveguides fabricated by direct bonding", Y. Shoji, T. Mizumoto, H. Yokoi, I-W. Hsieh, and R. M. Osgood, Appl. Phys. Lett. {\bf 92}, 071117 (2008).

\bibitem{Dillon2} "Magnetic and optical properties of rare earth garnets", J. F. Dillon, J. Magn. Magn. Mater. {\bf 84}, 213 (1990).

\bibitem{mcd-yig} "Magnetic circular dichroism and Faraday rotation spectra of Y$_{3}$Fe$_{5}$O$_{12}$", G. B. Scott, D. E. Lacklison, H. I. Ralph, and J. L. Page, Phys. Rev. B {\bf 12}, 2562 (1975).

\bibitem{Huard} S. Huard, \emph{Polarization of Light} (Wiley, 1997).

\bibitem{chip} "Silicon photonics: Lighting up the chip", A. Mekis, Nat. Photonics {\bf 2}, 389 (2008).

\bibitem{proximity} "Magnetostatic interaction in arrays of nanometric permalloy wires: A magneto-optic Kerr effect and a Brillouin light scattering study", G. Gubbiotti, S. Tacchi, G. Carlotti, P. Vavassori, N. Singh, S. Goolaup, and A. O. Adeyeye, Phys. Rev. B {\bf 72}, 224413 (2005).

\end{thebibliography}
\end{document}